\documentclass[12pt,preprint]{aastex}
\usepackage{epsfig}
\usepackage{natbib}
\usepackage{graphicx}
\usepackage{slashbox}
\usepackage{multirow}
\usepackage{lscape}
\usepackage{mathrsfs,amssymb}
\usepackage{amsmath}
\newcommand       \Angstrom     {\,{\rm \AA}}

\newcommand       \cm           {\,{\rm cm}}

\newcommand       \eV           {\,{\rm eV}}
\newcommand       \g            {\,{\rm g}}

\newcommand       \simlt        {\lesssim}

\newcommand       \mum          {\,{\rm \mu m}}
\newcommand       \ppm          {\,{\rm ppm}}

\newcommand       \simali       {\sim\,}

\newcommand	  \NC           {N_{\rm C}}
\newcommand	  \NH           {N_{\rm H}}
\newcommand	  \mre          {m^{\prime}}
\newcommand	  \mim          {m^{\prime\prime}}
\newcommand	  \epsre        {\varepsilon^{\prime}}
\newcommand	  \epsim        {\varepsilon^{\prime\prime}}
\newcommand	  \CTOHgraphene {\left[{\rm C/H}\right]_{\rm graph}}
\newcommand	  \GHz          {\,{\rm GHz}}
\def    \Nb	{M}
\def    \bT	{{\bf T}}

\def    \abs       {{\rm abs}}

\countdef\decade=200
\advance\decade by \year
\countdef\hours=201
\advance\hours by \time
\divide\hours by 60
\countdef\mins=202
\advance\mins by \hours
\multiply\mins by 60
\multiply\hours by 100
\countdef\miltime=203
\advance\miltime by \hours
\advance\miltime by \time
\advance\miltime by -\mins
\def\today{\number\decade.\number\month.\number\day.\number\miltime}
\shorttitle{Interstellar Graphene}

\begin{document}
\title{
\vspace*{-2.0em}
{\normalsize\rm Accepted for publication in
               {\it The Astrophysical Journal}}\\
\vspace*{1.0em}
On Graphene in the Interstellar Medium
\\{\small DRAFT: \today ~~}
}

\author{
X.H.~Chen\altaffilmark{1,2},
Aigen~Li\altaffilmark{2}, and
Ke~Zhang\altaffilmark{3}
}
\altaffiltext{1}{Department of Physics, Xiangtan University, 
                 411105 Xiangtan, Hunan Province, China
                }
\altaffiltext{2}{Department of Physics and Astronomy,
                 University of Missouri,
                 Columbia, MO 65211, USA;
                 {\sf lia@missouri.edu}
                 }
\altaffiltext{3}{Department of Astronomy,
                 University of Michigan,
                 Ann Arbor, MI 48109, USA
                }

\begin{abstract}
The possible detection of C$_{24}$, a planar graphene, 
recently reported in several planetary nebulae
by Garc{\'{\i}}a-Hern{\'a}ndez et al.\ (2011, 2012)
inspires us to explore 
whether and how much graphene could exist 
in the interstellar medium (ISM) and how 
it would reveal its presence through its 
ultraviolet (UV) extinction and 
infrared (IR) emission. 
In principle, interstellar graphene 
could arise from 
the photochemical processing of
polycyclic aromatic hydrocarbon (PAH) molecules 
which are abundant in the ISM through a complete 
loss of their hydrogen atoms, 
and/or from graphite which is thought to
be a major dust species in the ISM
through fragmentation caused by 
grain-grain collisional shattering. 
Both quantum-chemical computations 
and laboratory experiments have shown 
that the 
exciton-dominated
electronic transitions in graphene
cause a strong absorption band 
near 2755$\Angstrom$. 
%
We calculate the UV absorption of 
graphene and place an upper limit 
of $\simali$5$\ppm$ of C/H
(i.e., $\simali$1.9\% of 
the total interstellar C)
on the interstellar graphene abundance. 
We also model the stochastic heating 
of graphene C$_{24}$ in the ISM,
excited by single starlight photons 
of the interstellar radiation field
and calculate its IR emission spectra. 
We also derive the abundance of graphene 
in the ISM to be $<$\,5$\ppm$ of C/H 
by comparing the model emission spectra 
with that observed in the ISM.
\end{abstract}



\section{Introduction\label{intro}}
Carbon is exclusively formed 
in the hot interiors of stars 
through the fusion reactions of
three alpha particles (i.e., helium nuclei) 
and expelled into the interstellar space 
through stellar outflows
and/or supernova explosions 
in the late stages of stellar evolution.
As the fourth most abundant element 
in the universe (by mass) after hydrogen,
helium, and oxygen, carbon plays an important 
role in the physical and chemical evolution 
of the interstellar medium 
(ISM; Henning \& Salama 1998).
Due to its unique property to form three 
different types of chemical bonds through 
sp$^{1}$, sp$^{2}$, and sp$^{3}$ hybridizations, 
carbon can be stabilized in various multi-atomic 
structures with different molecular configurations 
called allotropes, including amorphous carbon, graphite, 
diamond, polycyclic aromatic hydrocarbon (PAH),
fullerenes, graphene, and carbon nanotubes. 

Many allotropes of carbon are known to be present 
in the ISM (e.g., see J\"ager et al.\ 2011). 
Presolar graphite grains and nanodiamonds
have been identified in primitive meteorites based 
on their isotopically anomalous composition.
These presolar grains were originally condensed 
in the cooling outflows 
from carbon-rich evolved stars
or in the ejecta of supernovae.
They were then expelled into the ISM 
and eventually made their way into 
the parental molecular cloud of the Sun
and incorporated into asteroids,  
the parent bodies of meteorites. 
Therefore, prior to their incorporation 
into asteroids during the early stages 
of solar system formation,
presolar grains must have had 
a sojourn in the ISM   
(see Li \& Mann 2012). 
While hydrogenated
amorphous carbon (HAC) grains reveal their 
presence in the diffuse ISM through the ubiquitous 
3.4$\mum$ aliphatic C--H absorption feature
(Pendleton \& Allamandola 2002),
the aromatic C--H and C--C emission features
at 3.3, 6.2, 7.7, 8.6 and 11.3$\mum$
infer the widespread presence of PAHs 
in a wide variety of interstellar regions
(e.g., see Tielens 2008).  
The detections of interstellar C$_{60}$ 
and C$_{70}$ and their cations have also 
been reported based on their characteristic
infrared (IR) emission features, 
e.g., the 7.0, 8.45, 17.3 and 18.9$\mum$ features 
of C$_{60}$ (Cami et al.\ 2010, 
Sellgren et al.\ 2010,  
Zhang \& Kwok 2011, Bern\'e et al.\ 2017,
Garc{\'{\i}}a-Hern{\'a}ndez et al.\ 2010)
and the 6.4, 7.1, 8.2 and 10.5$\mum$ features
of C$_{60}^{+}$ (Bern\'e et al.\ 2013,
Strelnikov et al.\ 2015). 
In addition,
C$_{60}^{+}$ has also been suggested as 
a promising carrier of the mysterious 
diffuse interstellar bands at 9348.4, 9365.2, 9427.8,
9577.0, and 9632.1$\Angstrom$
(Foing \& Ehrenfreund 1994,
Campbell et al.\ 2015, 2016, 
Walker et al.\ 2015).

Graphene was first synthesized in the laboratory 
in 2004 by A.K.~Geim and K.S.~Novoselov
(see Novoselov et al.\ 2004) for which they 
received the 2010 Nobel Prize in physics.\footnote{%
  We should note that, prior to the work of 
  A.K.~Geim and K.S.~Novoselov, the Jena 
  Laboratory Astrophsyics Group had already
  successfully synthesized graphene
  (e.g., see Henning et al.\ 2004).
  }
More recently, 
Garc{\'{\i}}a-Hern{\'a}ndez et al.\ (2011b, 2012)
reported for the first time the presence of
unusual IR emission features at $\simali$6.6, 9.8, 
and 20$\mum$ in several planetary nebulae (PNe), 
both in the Milky Way and the Magellanic Clouds,
which are coincident with the strongest transitions 
of planar C$_{24}$, a piece of graphene.
If confirmed,
this would be the first detection 
of graphene in space.
In principle, graphene could be present 
in the ISM as it could be formed from
the photochemical processing of PAHs,
which are abundant in the ISM,
through a complete loss of 
their H atoms (e.g., see Bern\'e \& Tielens 2012).
Chuvilin et al.\ (2010) showed experimentally 
that C$_{60}$ could be formed from a graphene sheet.
Bern\'e \& Tielens (2012) further proposed that
such a formation route could occur in space.
On the other hand, 
if there exists in the ISM 
a population of HAC-like nanoparticles
with a mixed aromatic/aliphatic structure
(e.g., see Kwok \& Zhang 2011), 
a complete loss of their H atoms 
could also convert HAC-like 
nanoparticles into graphene 
(e.g., see  Garc\'ia-Hern\'andez et al.\ 
2010, 2011a,b).
Also, graphene could be generated 
in the ISM from the exfoliation of graphite
as a result of grain-grain collisional fragmentation.
It is worth noting that graphite is thought to be a major 
dust component in the ISM (Draine \& Lee 1984)
and as mentioned earlier, presolar graphite grains 
have been identified in primitive meteorites. 
%


Both quantum-chemical computations 
and laboratory experiments have shown 
that the 
exciton-dominated
electronic transitions in graphene
cause a strong absorption band 
near 2755$\Angstrom$ 
(Yang et al.\ 2009, Nelson et al.\ 2010, 
Trevisanutto et al.\ 2010)
which is not seen in the ISM.
In this work, we aim at placing 
a quantitative upper limit on
the abundance of interstellar graphene
on the basis of the nondetection 
of the fingerprint 2755$\Angstrom$ 
absorption feature in the diffuse ISM.
To achieve this, we first calculate
in \S\ref{sec:extcurv} 
the UV absorption of graphene 
and compare it with the Galactic
interstellar extinction curve.
Also, if graphene is present in the diffuse ISM,
single-photon heating by starlight 
(Draine \& Li 2001)
will cause it to radiate in the IR
through its characteristic vibrational transitions
(e.g., see Garc{\'{\i}}a-Hern{\'a}ndez et al.\ 2011b,
Mackie et al.\ 2015).
Therefore, we calculate in \S\ref{sec:irem}
the IR emission spectrum of graphene 
heated by starlight, 
and compare it with measurements of 
the IR emission of the diffuse ISM 
by the {\it Infrared Telescope in Space}
(IRTS; Onaka et al.\ 1996) and by
the {\it Diffuse Infrared Background 
Experiment} (DIRBE; Arendt et al.\ 1998) 
and the {\it Far Infrared Absolute
Spectrophotometer} (FIRAS; 
Finkbeiner et al.\ 1999) instruments on 
the {\it Cosmic Background Explorer} 
(COBE) satellite.
The major conclusion is summarized 
in \S\ref{sec:summary}.

\section{UV Absorption}\label{sec:extcurv}
\begin{figure}[h]
\centering
\includegraphics[height=8cm,width=8cm]{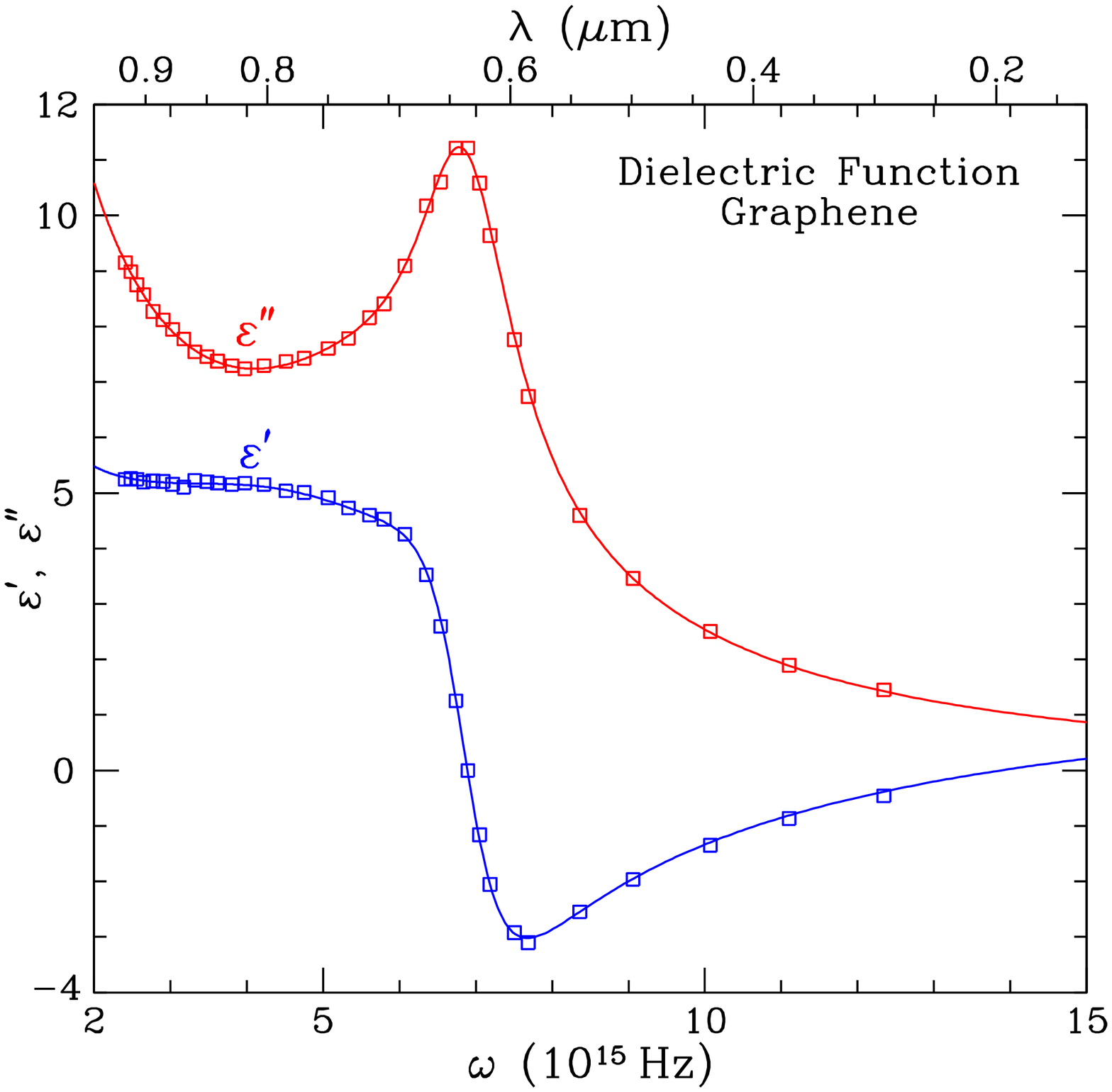}
\vspace{20mm}
\caption{\footnotesize
         \label{fig:nk}
         Real ($\epsre$) and imaginary ($\epsim$)
         parts of the dielectric function of graphene.
         Open squares: experimental data of 
         Nelson et al.\ (2010).
         Solid lines: model fits with three 
         Lorentz oscillators.   
         }
\end{figure}

For a planar graphene, the absorption cross section 
$C_{\rm abs}(\lambda)$ at wavelength $\lambda$ 
per unit volume ($V$) can be derived 
from the dielectric function $\varepsilon$ 
as follows (see Yang et al.\ 2009):
\begin{equation}\label{eq:Cabs2V}
C_{\rm abs}(\lambda)/V = 
\left(2\pi/\lambda\right)\,{\rm Im}\left\{\varepsilon-1\right\} ~~, 
\end{equation}
where ${\rm Im\left\{...\right\}}$ 
denotes the imaginary part of a complex function.
Nelson et al.\ (2010) employed spectroscopic 
ellipsometry to measure the complex
refractive indices, $m(\lambda) = \mre + i\,\mim$ 
of graphene in the wavelength range of 
$0.153 < \lambda <0.783\mum$
or $1.28 < \lambda^{-1} <6.54\mum^{-1}$.
We first convert the refractive indices of 
graphene measured by Nelson et al.\ (2010) 
to complex dielectric functions 
$\varepsilon = \epsre + i\,\epsim = m^2$
(see Figure~\ref{fig:nk}).
The Galactic interstellar extinction curve 
has been relatively well determined 
from the near-IR at $\lambda^{-1}\simlt1\mum^{-1}$ 
to the far-UV at $\lambda^{-1}\approx10\mum^{-1}$,
over a wavelength range {\it wider} than 
that of Nelson et al.\ (2010). 
To facilitate a direct comparison of
the absorption of graphene 
with the interstellar extinction curve, 
an extrapolation of the dielectric functions of
Nelson et al.\ (2010) over a wide wavelength range
is needed. To this end, we fit the dielectric functions 
of Nelson et al.\ (2010)
with three Lorentz oscillators
(see Bohren \& Huffman 1983):
\begin{equation}\label{eq:epsre}
\epsre(\omega) = \varepsilon_0 + \sum_{j=1}^{3}
\frac{\omega_{p,j}^2\left(\omega_{0,j}^2-\omega^2\right)}
{\left(\omega_{0,j}^2-\omega^2\right)^2
+ \gamma_j^2\omega^2} ~~,
\end{equation}
\begin{equation}\label{eq:epsim}
\epsim(\omega) = \sum_{j=1}^{3}
\frac{\omega_{p,j}^2\gamma_j\omega}
{\left(\omega_{0,j}^2-\omega^2\right)^2
+ \gamma_j^2\omega^2} ~~,
\end{equation}
where $\omega=2\pi c/\lambda$ 
is the angular frequency, 
$c$ is the speed of light,
$\omega_{p,j}$, $\omega_{0,j}$, and $\gamma_j$
are respectively the plasma frequency,
resonant frequency, and damping constant
of the $j$-th oscillator, and
$\varepsilon_0$ is the dielectric function 
at high frequencies 
(i.e., $\omega\gg\omega_0$).
As illustrated in Figure~\ref{fig:nk},
with $\varepsilon_0\approx1.71$
and three Lorentz oscillators 
characterized by
$\omega_{p,1}\approx7.79\times10^{15}\GHz$,
$\omega_{0,1}\approx6.89\times10^{15}\GHz$,
$\gamma_{1}\approx1.52\times10^{15}\GHz$,
$\omega_{p,2}\approx16.47\times10^{15}\GHz$,
$\omega_{0,2}\approx1.61\times10^{15}\GHz$,
$\gamma_{2}\approx13.99\times10^{15}\GHz$,  
$\omega_{p,3}\approx10.58\times10^{15}\GHz$,
$\omega_{0,3}\approx6.33\times10^{15}\GHz$,
$\gamma_{3}\approx5.18\times10^{15}\GHz$,
we are able to closely fit the dielectric 
functions of graphene experimentally derived
by Nelson et al.\ (2010).
Eqs.\,\ref{eq:epsre} and \ref{eq:epsim} allow us
to extrapolate the dielectric functions of graphene
at $\lambda^{-1} < 1.28\mum^{-1}$
and $\lambda^{-1} > 6.54\mum^{-1}$.\footnote{%
   This extrapolation is physical 
   since it is based on a simple physical principle
   (i.e.,  the classical Lorentz harmonic oscillator model).
   Even if this extrapolation is not accurate,
   it does not affect the following discussion
   on its contribution to the interstellar extinction
   since the absorption of graphene occurs mostly
   at $2 < \lambda^{-1} <6\mum^{-1}$. 
   Also because of this, it will not affect
   the following discussion on its IR emission.
   After all, graphene absorbs little  
   at $\lambda^{-1} < 1.28\mum^{-1}$
   and $\lambda^{-1} > 6.54\mum^{-1}$
   and therefore the existing, experimentally-derived
   dielectric function data 
   at $1.28 < \lambda^{-1} <6.54\mum^{-1}$
   are sufficient for determining the heating 
   of graphene.
   }

For graphene, we can relate its volume ($V$) 
to the number of C atoms ($\NC$), through
\begin{equation}\label{eq:NC}
\frac{V}{\NC} = \frac{12\,m_{\rm H}\,d}{\sigma} ~~,
\end{equation}
where $m_{\rm H}\approx1.66\times10^{-24}\g$ 
is the mass of a hydrogen nuclei,
$d\approx3.4\Angstrom$ is the monolayer
thickness of graphene, and 
$\sigma\approx6.5\times10^{-8}\g\cm^{-2}$
is the surface mass density of graphene.
Combining eq.\,\ref{eq:Cabs2V} 
with eq.\,\ref{eq:NC}, we obtain the absorption
cross section of graphene per C atom from
\begin{equation}\label{eq:Cabs2NC}
C_{\rm abs}(\lambda)/\NC = \frac{24\pi\,m_{\rm H}\,d}
{\sigma\lambda}\,{\rm Im}\left\{\varepsilon-1\right\} ~~.
\end{equation}

\begin{figure}[h]
\centering
\includegraphics[height=8cm,width=8cm]{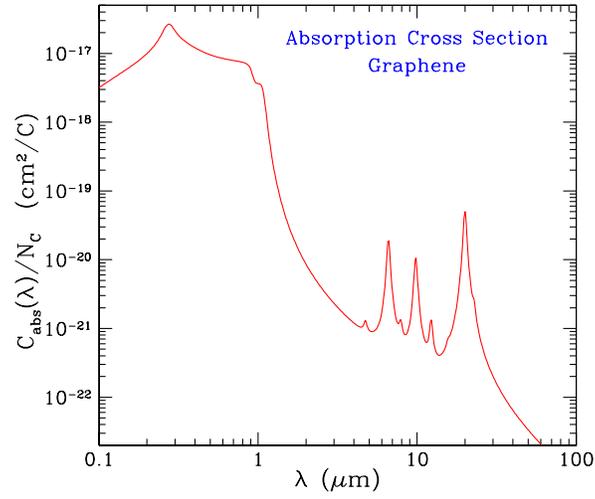}
\vspace{20mm}
\caption{\footnotesize
         \label{fig:Cabs}
         Absorption cross sections (per C atom)
         of graphene from the far-UV to the IR.
         Following Li \& Draine (2001b), 
         we smoothly join the UV part and the IR part.
         }
\end{figure}

\begin{figure}[h]
\centering
\includegraphics[width=.5\textwidth]{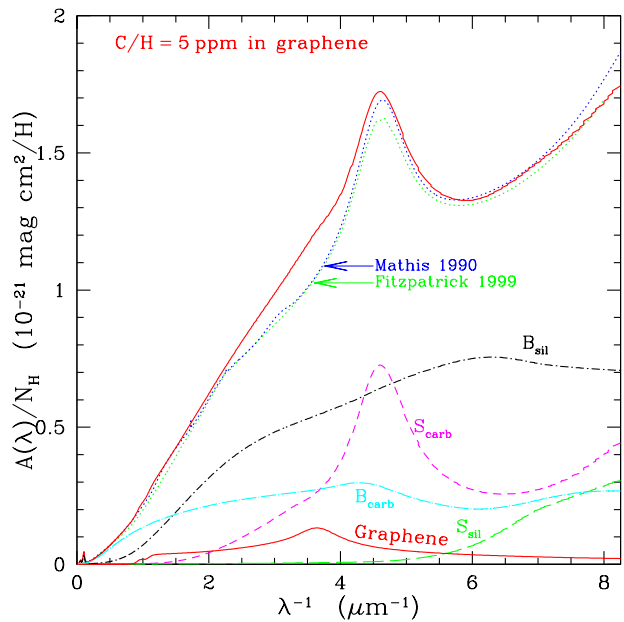}
\vspace{20mm}
\caption{\footnotesize
         \label{fig:extcurv}
        Comparison of the average Galactic interstellar extinction 
        curve (open circles: Mathis 1990; dotted line: Fitzpatrick 
        1999) with the model extinction curve (solid red line)
        obtained by adding the contribution from graphene
        with $\CTOHgraphene=5\ppm$ (thin red line) 
        to the best-fit model of Weingartner \& Draine (2001).
        Also plotted are the contributions 
        (see Li \& Draine 2001b) from 
        ``${\rm B_{sil}}$'' ($a \ge 250$\AA\ silicate);
        ``${\rm S_{sil}}$'' ($a < 250$\AA\ silicate); 
        ``${\rm B_{carb}}$'' ($a\ge 250$\AA\ carbonaceous); 
        ``${\rm S_{carb}}$'' ($a < 250$\AA\ carbonaceous, 
        including PAHs).       
        }
\end{figure}

In Figure~\ref{fig:Cabs} we show the UV/optical 
absorption cross section (per C atom) of graphene.
Most prominent in the absorption profile of graphene
is the exciton-dominated absorption peak 
at $\simali$4.5$\eV$ (i.e, $\lambda\approx2755\Angstrom$,
$\lambda^{-1}\approx3.63\mum^{-1}$).
We note that quantum-chemical first-principles 
calculations of many-electron effects on 
the optical response of graphene have also
demonstrated that the resonant excitons 
give rise to a prominent peak 
in the absorption spectrum of graphene
near 4.5$\eV$ (see Yang et al.\ 2009,
Trevisanutto et al.\ 2010).
However, this absorption peak is absent 
in the interstellar extinction curve. 
As shown in Figure~\ref{fig:extcurv},
the Galactic extinction curve 
--- the variation of the extinction $A_\lambda$
with the inverse wavelength $\lambda^{-1}$
--- rises almost linearly from the near-IR 
to the near-UV, with a broad absorption bump 
at about $\lambda^{-1}$$\approx$4.6$\mum^{-1}$
($\lambda$$\approx$2175$\Angstrom$)
and followed by a steep rise into the far-UV
at $\lambda^{-1}$$\approx$10$\mum^{-1}$,
the shortest wavelength at which 
the dust extinction has been measured
(Mathis 1990, Fitzpatrick 1999).
The nondetection of the 2755$\Angstrom$ 
absorption feature allows us to place an 
upper limit on the abundance of 
graphene in the ISM.

Weingartner \& Draine (2001) 
and Li \& Draine (2001b) have developed
a carbonaceous-silicate grain model 
which reproduces both the observed Galactic 
extinction curve and the observed IR emission.
In Figure~\ref{fig:extcurv}
we show the extinction obtained 
by {\it adding} graphene to this model.
Since graphene is in the Rayleigh limit, 
the added extinction depends only on 
$\CTOHgraphene$, the amount of C 
(relative to H) tied up in graphene, 
and is independent of the exact graphene size:
\begin{equation}
\left(\frac{A_\lambda}{\NH}\right)_{\rm graph}
= 1.086\,\left(\frac{C_{\rm abs}}{\NC}\right)_{\rm graph}
\CTOHgraphene ~~.
\end{equation}

As shown in Figure~\ref{fig:extcurv}, 
the red wing of the strong 2175$\Angstrom$ 
extinction bump could hide certain amount 
of graphene: we estimate the upper bound to 
be $\CTOHgraphene\approx5\ppm$ ---
i.e., in the diffuse ISM there could 
be as much as $\simali$5$\ppm$ of C/H in graphene
for the characteristic 2755$\Angstrom$
absorption feature of graphene 
to remain unnoticeable. 
The contribution of graphene 
to the $\lambda^{-1}\sim3.6\mum^{-1}$ region 
would be considerable for $\CTOHgraphene>6\ppm$.
If we take the interstellar C abundance 
to be solar-like
(i.e., C/H\,$\approx$\,269$\pm$31$\ppm$,
Asplund et al.\ 2009), an upper limit of
 $\CTOHgraphene\approx5\ppm$
implies that at as much as $\simali$1.9\% of 
the interstellar C atoms 
could be tied up in graphene.

\section{IR Emission}\label{sec:irem}
For a planar graphene of several hundred 
C atoms or smaller,
photon absorption is often 
the dominant excitation process
and its energy content is often smaller than
the energy of a single starlight photon.  
Therefore, graphene will undergo stochastic
heating in the ISM (Greenberg 1968).
In most cases soon after 
the photoabsorption, graphene will 
convert almost all of the initial 
photoexcitation energy to vibrational energy 
of the highly vibrationally 
excited ground electronic state, 
and hence IR emission is always 
the dominant relaxation process
(e.g., see Li 2004).
Therefore, we will model the stochastic heating 
of graphene in terms of pure vibrational transitions
by employing the ``exact-statistical'' method
of Draine \& Li (2001).

Following Draine \& Li (2001), 
we characterize the state of 
a graphene sheet
of $\NC$ C atoms by its vibrational energy $E$, 
and group its energy levels into 
$(\Nb+1)$ ``bins'',
where the $j$-th bin ($j$\,=\,0,...,$\Nb$) is 
$[E_{j,\min},E_{j,\max})$, 
with representative energy 
$E_j$\,$\equiv$\,$(E_{j,\min}$+$E_{j,\max})$/2,
and width 
$\Delta E_j$\,$\equiv$\,$(E_{j,\max}$--$E_{j,\min})$.
Let $P_j$ be the probability of finding 
the graphene sheet of $\NC$ C atoms 
in bin $j$ with energy $E_j$.
The probability vector $P_j$ evolves 
according to
\begin{equation}
dP_i/dt = \sum_{j\neq i} \bT_{ij} P_j 
- \sum_{j\neq i} \bT_{ji}P_i ~~,~~ i\,=\,0,...,\Nb ~~,
\end{equation}
where the transition matrix element $\bT_{ij}$ is
the probability per unit time for
graphene in bin $j$ 
to make a transition to one of the levels in bin $i$. 
We solve the steady state equations
\begin{equation}
\sum_{j\neq i} \bT_{ij} P_j
= \sum_{j\neq i} \bT_{ji}P_i ~~,~~ i\,=\,0,...,\Nb ~~
\end{equation}
to obtain the $\Nb$+1 elements of $P_j$,
and then calculate the resulting IR emission spectrum
(see eq.\,55 of Draine \& Li 2001).

Before we proceed to calculate the state-to-state 
transition rates $\bT_{ji}$ for transitions 
$i$$\rightarrow$$j$, lets distinguish 
the excitation rates $\bT_{ul}$ 
(from $l$ to $u$, $l$\,$<$\,$u$) 
from the deexcitation rates $\bT_{lu}$ 
(from $u$ to $l$, $l$\,$<$\,$u$).
For a given starlight energy density $u_E$,
the rates for upward transitions $l$$\rightarrow$$u$ 
(i.e., the excitation rates)
are just the photon absorption rates:
\begin{equation}
\bT_{ul} \approx C_{\abs}(E)\,c\,u_E \Delta E_u/(E_u-E_l) ~~.
\end{equation}
The rates for downward transitions $u$$\rightarrow$$l$
(i.e., the deexcitation rates)
can be determined from the detailed balance analysis
of the Einstein $A$ coefficient:
\begin{equation}
\bT_{lu} \approx \frac{8\pi}{h^3c^2} \frac{g_l}{g_u}
\frac{\Delta E_u}{E_u-E_l} E^3 \times C_{\abs}(E) 
\left[1+\frac{h^3c^3}{8\pi E^3}u_E\right] ~~,
\end{equation}
where $h$ is the Planck constant,
and the degeneracies $g_u$ and $g_l$ 
are the numbers of energy states 
in bins $u$ and $l$, respectively:
\begin{equation}
\label{eq:gj}
g_j \equiv N(E_{j,\max})-N(E_{j,\min})
\approx \left(dN/dE\right)_{E_j} \Delta E_j ~~,
\end{equation}
where $\left(dN/dE\right)_{E_j}$
is the vibrational density of states 
of graphene at internal energy $E_j$.
For a planar graphene sheet of $\NC$ C atoms,
if we know the frequencies of 
all its $\left(3\NC-6\right)$ vibrational modes,
we can employ the Beyer-Swinehart numerical algorithm
(Beyer \& Swinehart 1973, Stein \& Rabinovitch 1973)
to calculate the vibrational density of states 
$\left(dN/dE\right)_{E_j}$
and therefore the degeneracies $g_j$ 
for each vibrational energy bin.
If we know the oscillator strength
of each vibrational mode, 
we can obtain the IR absorption
cross section $C_{\rm abs}(E)$ of graphene
by summing up all the vibrational transitions
with each approximated as a Drude profile.

Mackie et al.\ (2015) computed the vibrational
transitions of 805 fully dehydrogenated PAHs,
using the B3LYP density functional theory (DFT),
together with the 4-31G basis set (Frisch et al.\ 1984). 
These molecules span a wide range of sizes,
from two benzene rings up to eight rings,
with the largest one containing 34 C atoms.
Although the individual spectra of these fully 
dehydrogenated species are rather diverse,
as a whole they do show characteristic features 
at 5.2, 5.5, and 10.6$\mum$ as well as a forest 
of features in the $\simali$16--30$\mum$ range 
that appears as a structured continuum, but with 
a clear peak centered around 19$\mum$. 
However, we note that not all of these fully 
dehydrogenated PAHs are planar or graphene-like.
Also, a complete set of the frequencies of 
all the $\left(3\NC-6\right)$ vibrational modes
computed by Mackie et al.\ (2015) 
for each individual species of $\NC$ C atoms 
is not yet publicly available. 
This prevents us from computing the IR emission
spectra of the graphene-like species considered
by Mackie et al.\ (2015). 
This is because,  
to calculate the vibrational density of states 
(see eq.\,\ref{eq:gj}),
for a given species of $\NC$ C atoms
a complete knowledge of the frequencies 
of all the $\left(3\NC-6\right)$ 
vibrational modes is required.
On the other hand, 
in the literature
the vibrational modes and intensities
are known for C$_{24}$, 
a small graphene sheet.
Martin et al.\ (1996) performed
quantum-chemical computations 
for C$_{24}$ using the B3LYP DFT.
Kuzmin \& Duley (2011) also
carried out DFT-based calculations
for planar C$_{24}$ and obtained
similar results.
In this work, we will therefore
consider the vibrational excitation
and the subsequent IR emission of C$_{24}$.
We note that C$_{24}$ is preferred 
also because it has been possibly 
detected in several Galactic 
and extragalactic PNe
(Garc{\'{\i}}a-Hern{\'a}ndez et al.\ 2011a, 2012)
and even in the ISM (Bern\'e et al.\ 2013).\footnote{%
    Bern\'e et al.\ (2013) reported the detection
    in NGC\,7023, a reflection nebula, 
    of the 6.6$\mum$ band coincident with 
    the assignment of C$_{24}$ made by 
    Garc{\'{\i}}a-Hern{\'a}ndez et al.\ (2011a).
    }

We consider the stochastic heating of
C$_{24}$ in the diffuse ISM, excited
by the solar neighbourhood interstellar
radiation field (ISRF) estimated by 
Mathis et al.\ (1983; hereafter MMP83).
We consider 500 energy bins
(i.e., $M=500$).
We use the vibrational modes 
and intensities of C$_{24}$
computed by Martin et al.\ (1996) 
to calculate the vibrational density of states 
and the degeneracies 
for each vibrational energy bin
as well as the IR absorption cross section.
Without a prior knowledge of the width
of each vibrational transition, 
we assign a width of 30$\cm^{-1}$, 
consistent with the natural line width 
expected from free-flying molecule 
(see Allamandola et al.\ 1989).
This natural line width arises from 
intramolecular vibrational energy redistribution. 
In Figure~\ref{fig:Cabs} we show 
the IR absorption cross section 
of graphene based on the quantum-chemical
data of C$_{24}$ which clearly exhibits
three strong IR bands at 6.6, 9.8 and 20$\mum$.

\begin{figure}[h]
\centering
\includegraphics[width=.5\textwidth]{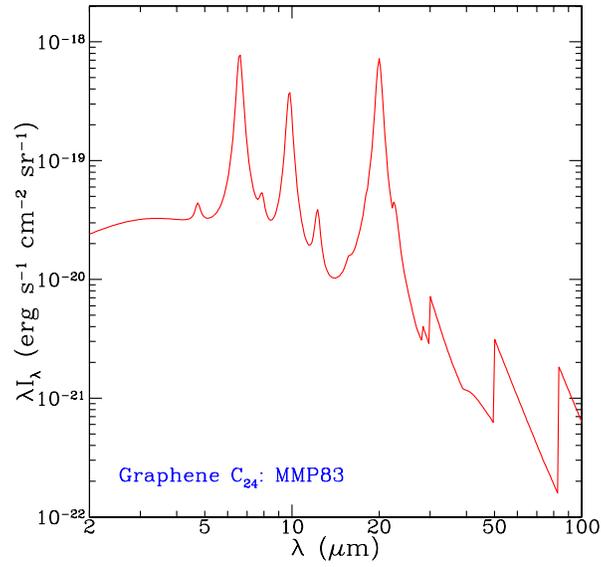}
\vspace{20mm}
\caption{\footnotesize
         \label{fig:irem}
         IR emission spectrum 
         of graphene of $\NC=24$
         illuminated by the MMP83 ISRF.
         The sawtooth features 
         at $\lambda>30\mum$ are
         due to our treatment of 
         transitions from the lower 
         excited energy bins to 
         the ground state and first 
         few excited states
         (see Draine \& Li 2001).
         }
\end{figure}

With $g_j$ and $C_{\rm abs}(E)$ 
derived from C$_{24}$ of Martin et al.\ (1996)
and the starlight energy density $u_E$
taken from MMP83,
we determine the state-to-state 
transition rates $\bT_{ji}$ 
and solve the steady-state probability 
evolution equation
to obtain the steady-state energy probability 
distribution $P_j$ 
and finally calculate 
the resulting IR emission spectrum.
In Figure~\ref{fig:irem} we show
the IR emission of C$_{24}$ excited
by the MMP83 ISRF.
Most pronounced are the IR emission features
at 6.6, 9.8 and 20$\mum$.
We note that the sawtooth features 
at $\lambda>30\mum$ are artificial; 
they are due to our treatment of 
transitions from the lower excited energy bins 
to the ground state and first few excited states
(see Draine \& Li 2001).

\begin{figure}[h]
\centering
\includegraphics[width=.5\textwidth, angle=270]{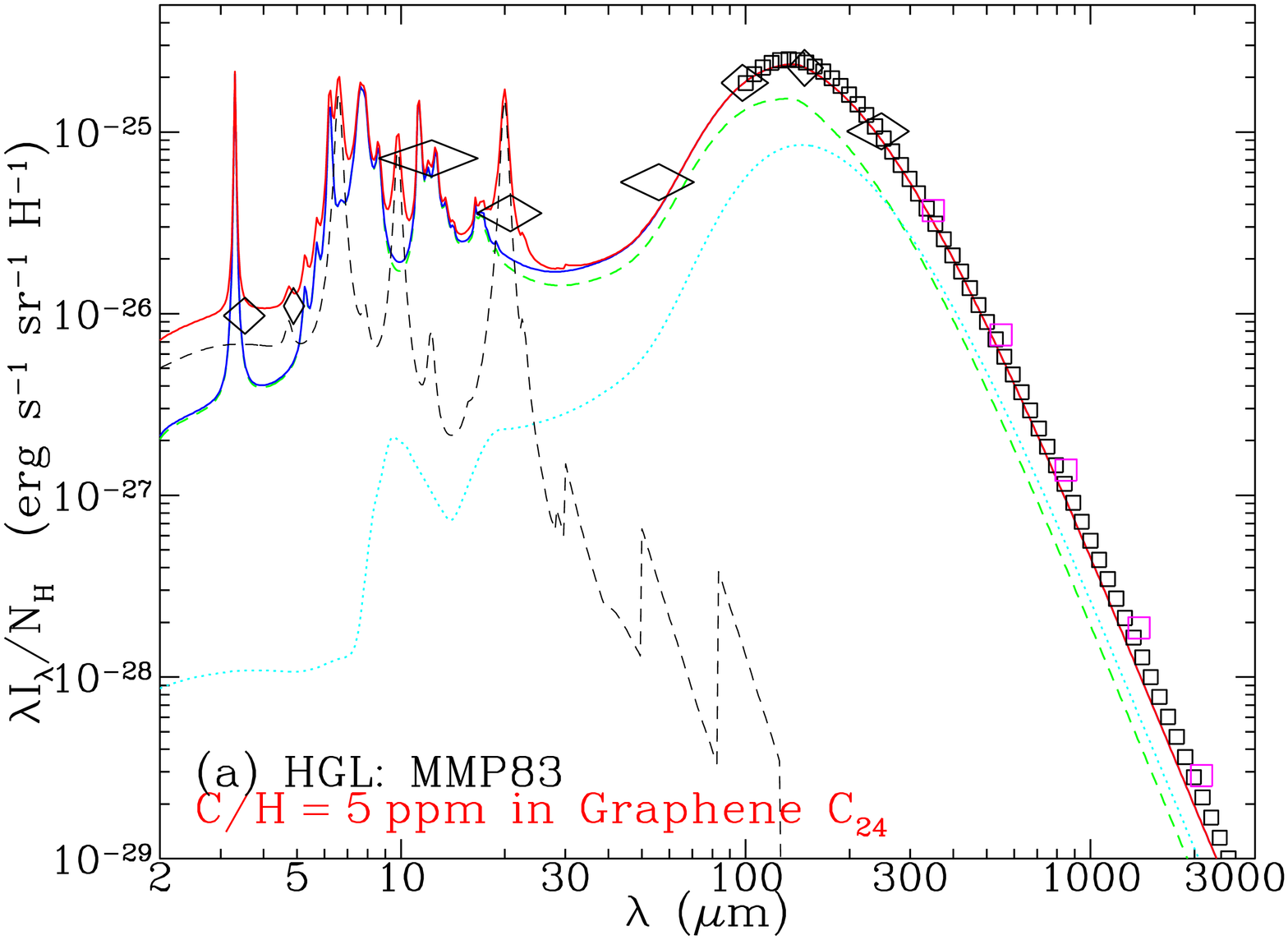}
\includegraphics[width=.5\textwidth, angle=270]{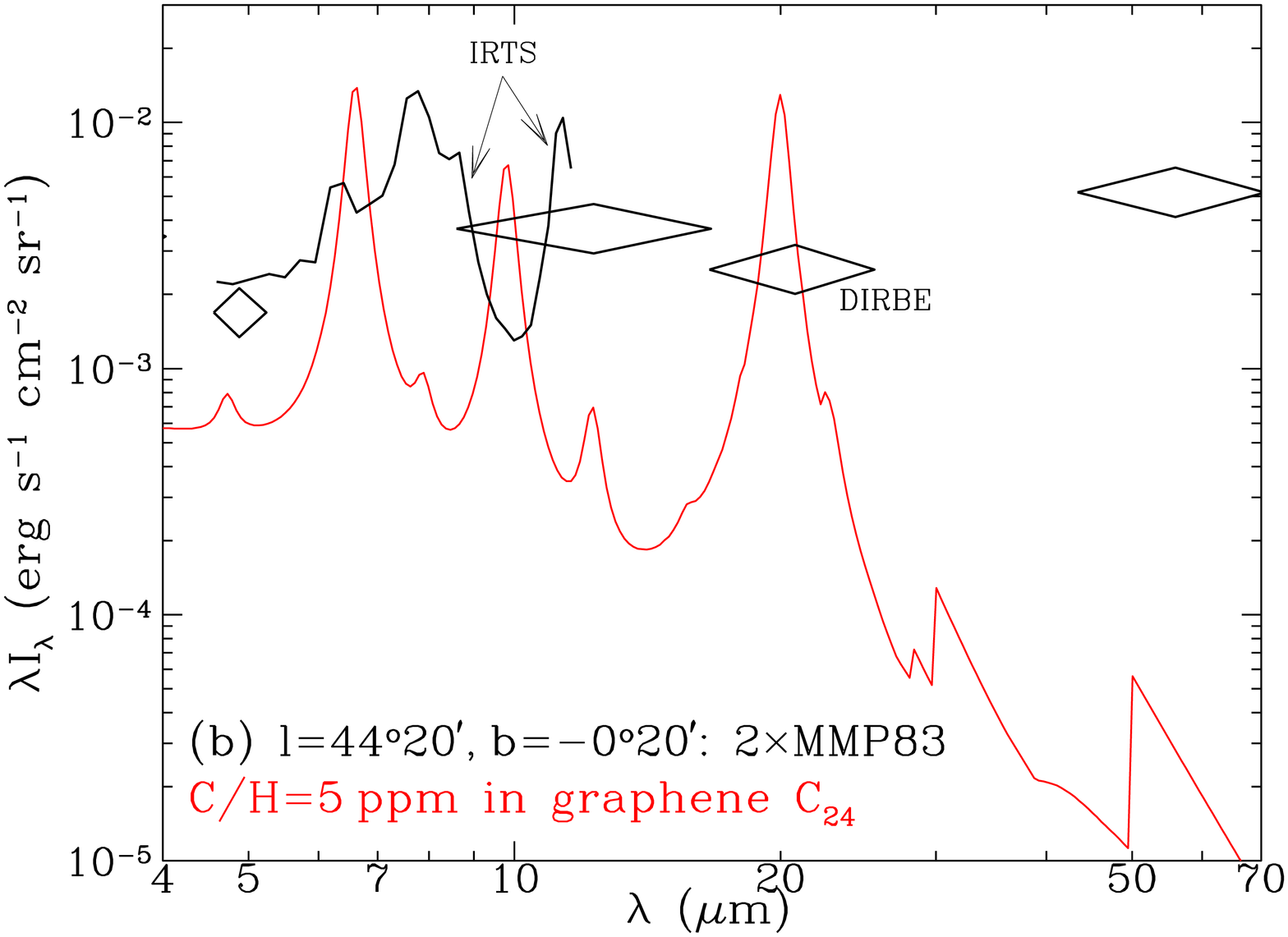}
\caption{\footnotesize
         \label{fig:dism}
        Upper panel (a):
        Comparison of the observed IR emission 
        from the HGL cirrus
        with the model emission spectrum 
        obtained by adding the contribution 
        from graphene C$_{24}$
        of $\CTOHgraphene=5\ppm$ (black dashed line) 
        to the best-fit model of Li \& Draine (2001b)
        which consists of contributions from
        amorphous silicate (cyan dotted line) and
        carbonaceous grains (i.e., graphite and PAHs;
        green dashed line). The sum of amorphous silicate 
        and carbonaceous grains is shown as blue solid line.
        Observational data are from 
        {\it DIRBE} (black diamonds; Arendt et al.\ 1998), 
        {\it FIRAS} (black squares; Finkbeiner et al.\ 1999), 
        and {\it Planck} (magenta squares; 
        Planck Collaboration XVII 2014).
        Bottom panel (b):
        Contribution to the IR emission toward
        ($44^{\rm o}\le l \le 44^{\rm o}40^\prime$, 
        $-0^{\rm o}40^\prime\le b \le 0^{\rm o}$) by
        graphene C$_{24}$
        of $\CTOHgraphene=5\ppm$ (solid red line). 
	Diamonds: {\it DIRBE} photometry.
	Black solid line:
        5--12$\mum$ spectrum observed by IRTS
        (Onaka et al.\ 1996). 
        }
\end{figure}

We will now derive upper limits on 
the abundance of graphene in the diffuse ISM 
based on comparison of the observed 
IR emission with the calculated emission 
spectrum of C$_{24}$.
We first consider the high Galactic-latitude (HGL) 
cirrus illuminated by the MMP83 ISRF.
The average emission per H for the HGL region
has been measured by {\it COBE}/DIRBE 
(Arendt et al.\ 1998), {\it COBE/FIRAS} 
(Finkbeiner et al.\ 1999), and {\it Planck} 
(Planck Collaboration XVII 2014). 
As shown in Figure~\ref{fig:dism}a,
with 5$\ppm$ of C/H in graphene C$_{24}$,
the 6.6, 9.8 and 20$\mum$ emission features
of graphene will become so prominent that
they will be essentially as strong as 
the PAH features at 6.2, 7.7, 8.6 and 11.3$\mum$ 
and would have been detected by {\it Spitzer} or 
by the {\it Short Wavelength Spectrometer} (SWS)
aboard the {\it Infrared Space Observatory} (ISO). 

Following Li \& Draine (2001a), 
we have also considered
the diffuse ISM toward
$l = 44^{\rm o}20^\prime$, 
$b=-0^{\rm o}20^\prime$ 
which has been observed
by {\it COBE/DIRBE}
(Hauser et al.\ 1998)
and the {\it Mid-Infrared Spectrograph} (MIRS) 
aboard {\it IRTS} 
has obtained the 4.7--11.7$\mu$m spectrum
(Onaka et al.\ 1996). 
Li \& Draine (2001b) have modeled 
the IR emission from the dust in this direction
and derived a total gas column 
$N_{\rm H}\approx4.3\times10^{22}\cm^{-2}$
and a starlight intensity of 
$U\approx2$ (in unit of the MMP83 ISRF).
As illustrated in Figure~\ref{fig:dism}b,
if graphene C$_{24}$ is present in
this region, its abundance must be less  
than 5$\ppm$ of C/H,
otherwise the 6.6, 9.8 and 20$\mum$ emission features
of graphene would have been detected by {\it IRTS}.
Therefore, for both the HGL region 
and the line of sight toward 
$l = 44^{\rm o}20^\prime$, $b=-0^{\rm o}20^\prime$,
a generous upper limit of C/H\,$\simlt$\,5$\ppm$
is imposed by the {\it COBE/DIRBE} photometric data
and the {\it IRTS} spectrum. 

\section{Summary}\label{sec:summary}
Inspired by the possible detection of
the 6.6, 9.8 and 20$\mum$ emission features
of graphene C$_{24}$ in several Galactic 
and extragalactic PNe, we have explored 
how much graphene could be present
in the diffuse ISM and how it would reveal 
its presence through absorption and emission.
We have placed an upper limit
of $\simali$5$\ppm$ of C/H on the abundance
of graphene in the diffuse ISM, 
based on the nondetection 
in the Galactic interstellar extinction curve 
of the prominent absorption peak 
at $\simali$2755$\Angstrom$ of graphene
caused by resonant excitons 
as well as the nondetection of
the 6.6, 9.8, and 20$\mum$ emission features
of graphene C$_{24}$ in the observed IR 
emission spectra of the diffuse ISM. 
While the extinction-based constraint is
size-independent as long as graphene is 
in the Rayleigh regime,
the IR-emission-based constraint is
sensitive to the graphene size
since the stochastic heating 
and the resulting IR emission spectrum
depend on the graphene size.
Further quantum-chemical computations 
and experimental measurements
of the IR vibrational spectra of graphene
of a wide range of sizes as well as
the vibrational excitation modeling
of the IR emission of graphene in
various astrophysical regions will
be crucial to more rigorously explore 
the presence and quantity of graphene in space.

\acknowledgments{
We thank B.T.~Draine, W.W.~Duley,
D.A.~Garc{\'{\i}}a-Hern{\'a}ndez,
Th.~Hening, C.~J\"ager, S.~Kuzmin, 
F.Y.~Xiang, L.~Yang, X.J.~Yang, 
J.X.~Zhong, and the anonymous referee
for very helpful discussions and suggestions.
This work is supported by NSFC through
Projects 11173007, 11373015, 11533002,
and 973 Program 2014CB845702.
AL is supported in part by
NSF AST-1311804 and NASA NNX14AF68G.
}

\clearpage

\clearpage

\clearpage

\clearpage

\clearpage

\begin{thebibliography}{}
\bibitem[]{}Allamandola, L.J., Tielens, A.G.G.M., 
                  \& Barker, J.R.\ 1989,
                  ApJS, 71, 733
 \bibitem[]{}Arendt, R. G., Odegard, N., Weiland, J. L., 
                   et al.\ 1998, ApJ, 508, 74
\bibitem[]{}Asplund, M., Grevesse, N., Sauval, A. J., 
                 \& Scott, P.\  2009, ARA\&A, 47, 481 
\bibitem[]{}Bern\'e, O, \& Tielens, A.G.G.M.\
                  2012, PNAS, 109, 401
\bibitem[]{}Bern{\'e}, O., Mulas, G., 
                  \& Joblin, C.\ 2013, 
                  A\&A, 550, L4 
\bibitem[]{}Bern{\'e}, O., Cox, N.~L.~J., 
                  Mulas, G., \& Joblin, C.\ 
                  2017, A\&A, 605, L1 
\bibitem[]{}Bernstein, L.~S., Shroll, R.~M., 
                  Lynch, D.~K., \& Clark, F.~O.\ 
                  2017, ApJ, 836, 229 
\bibitem[]{}Beyer, T., \& Swinehart, D.F.\ 1973,
                  Commun. Assoc. Comput. Machinery, 
                  16, 379
\bibitem[]{}Bohren, C.F., \& Huffman, D.R.\ 1983, 
                 Absorption and Scattering of Light by
                 Small Particles, Wiley, New York
\bibitem[]{}Campbell, E.K., Holz, M., Gerlich, D., 
                  \& Maier, J.P.\ 2015, 
                  Nature, 523, 322
\bibitem[]{}Campbell, E.K., Holz, M., \& Maier, J.P.\ 
                  2016, ApJL, 826, L4
\bibitem[]{}Cami, J., Bernard-Salas, J., Peeters, E., 
                  \& Malek, S.~E.\ 2010, Science, 329, 1180 
\bibitem[]{}Chuvilin, A., Kaiser, U., Bichoutskaia, E., 
                  Besley, N.~A., \& Khlobystov, A.~N.\ 
                  2010, Nature Chem., 2, 450 
\bibitem[]{}Draine, B.T., \& Lee, H.M.\
                 1984, ApJ, 285, 89
\bibitem[]{}Draine, B.T., \& Li, A.\ 
                 2001, ApJ, 551, 807 
\bibitem[]{}Finkbeiner, D. P., Davis, M., \&
                 Schlegel, D.J.\ 1999, ApJ, 524, 867
\bibitem[]{}Fitzpatrick, E.L.\ 1999, PASP, 111, 63
\bibitem[]{}Garc{\'{\i}}a-Hern{\'a}ndez, D.~A., 
                  Manchado, A., Garc{\'{\i}}a-Lario, P., 
                  et al.\ 2010, ApJL, 724, L39 
\bibitem[]{}Garc{\'{\i}}a-Hern{\'a}ndez, D.~A., 
                  Kameswara Rao, N., \& Lambert, D.~L.\ 
                  2011a, ApJ, 729, 126 
\bibitem[]{}Garc{\'{\i}}a-Hern{\'a}ndez, D.~A., 
                   Iglesias-Groth, S., Acosta-Pulido, J.~A., 
                   et al.\ 2011b, ApJL, 737, L30 
\bibitem[]{}Garc{\'{\i}}a-Hern{\'a}ndez, D.~A., 
                  Villaver, E., Garc{\'{\i}}a-Lario, P., 
                  et al.\ 2012, ApJ, 760, 107 
\bibitem[]{}Greenberg, J.M.\ 1968, 
            in Stars and Stellar Systems, Vol. VII, 
            ed. B.M. Middlehurst \& L.H. Aller 
            (Chicago: Univ. of Chicago Press), 221
\bibitem[]{}Hauser, M.G., Kelsall, T., Leisawitz, D., 
            \& Weiland, J.\ 1998,
            COBE Diffuse Infrared Background Experiment 
            Explanatory Supplement version 2.3
            (COBE Ref. Pub. No. 98-A; 
             Greenbelt: NASA/GSFC)
\bibitem[]{}Henning, Th., \& Salama, F.\ 
                 1998, Science, 282, 2204 
\bibitem[]{}Henning, Th., J{\"a}ger, C., 
                  \& Mutschke, H.\ 2004, 
                  in Astrophysics of Dust 
                  (ASP Conf. Ser. 309), 
                  ed. A.N. Witt, G.C. Clayton, 
                 \& B.T. Draine 
                 (San Francisco, CA: ASP), 603
\bibitem[]{}J{\"a}ger, C., Mutschke, H., Henning, Th., 
                 \& Huisken, F.\ 2011, 
                 EAS Publ. Ser., 46, 293 
\bibitem[]{}Kuzmin, S., \& Duley, W.~W.\ 
                 2011, arXiv:1103.2989
\bibitem[]{}Kwok, S., \& Zhang, Y.\ 2011, 
                 Nature, 479, 80 
\bibitem[]{}Li, A.\ 2004, Astrophysics of Dust 
                 (ASP Conf. Ser. 309), ed. A.N. Witt, 
                 G.C. Clayton, \& B.T. Draine 
                 (San Francisco, CA: ASP), 417
\bibitem[]{}Li, A., \& Draine, B.T.\ 
            2001a, ApJL, 550, L213
\bibitem[]{}Li, A., \& Draine, B.T.\ 
            2001b, ApJ, 554, 778
\bibitem[]{}Li, A., \& Mann, I.\ 2012,
            in Astrophys. Space Sci. Library, 
            Vol. 385, Nanodust in the Solar System: 
            Discoveries and Interpretations 
            (Berlin: Springer-Verlag), 5
\bibitem[]{}Mackie, C.~J., Peeters, E., 
                  Bauschlicher, C.~W., Jr., 
                  \& Cami, J.\ 2015, ApJ, 799, 131 
\bibitem[]{}Martin, J.~M.~L., El-Yazal, J., 
            \& Fran{\c c}ois, J.-P.\ 1996, 
            Chem. Phys. Lett., 255, 7 
\bibitem[]{}Mathis, J.S. 1990, ARA\&A, 28, 37
\bibitem[]{}Mathis, J.S., Mezger, P.G., \& Panagia, N. 
            1983, A\&A, 128, 212
\bibitem[]{}Nelson, F.~J., Kamineni, V.~K., Zhang, T., 
            et al.\ 2010, Appl. Phys. Lett., 97, 253110 
\bibitem[]{}Novoselov, K.~S., Geim, A.~K., Morozov, S.~V., 
            et al.\ 2004, Science, 306, 666 
\bibitem[]{}Onaka, T., Yamamura, I., Tanabe, T., 
            et al.\ 1996, PASJ, 48, L59
\bibitem[]{}Pendleton, Y.~J., \& Allamandola, L.~J.\ 
            2002, ApJS, 138, 75 
\bibitem[]{}Planck Collaboration XVII 2014,
            A\&A, 566, A55
\bibitem[]{}Sellgren, K., Werner, M.~W., 
            Ingalls, J.~G., et al.\ 2010, 
            ApJL, 722, L54 
\bibitem[]{}Stecher, T.P., \& Donn, B.\
            1965, ApJ, 142, 1681
\bibitem[]{}Stein, S.E., \& Rabinovitch, B.S.\ 
            1973, J. Chem. Phys., 58, 2438
\bibitem[]{}Strelnikov, D., Kern, B., 
            \& Kappes, M.~M.\ 2015, 
            A\&A, 584, A55 
\bibitem[]{}Tielens, A.~G.~G.~M.\ 2008, 
            ARA\&A, 46, 289 
\bibitem[]{}Trevisanutto, P.E., Holzmann, M.,
            C\^{o}t\'{e}, M., \& Olevano, V.
            2010, Phys. Rev. B, 81, 121405
\bibitem[]{}Walker, G.A.H., Bohlender, D. A., 
                  Maier, J. P., \& Campbell, E. K.\
                  2015, ApJL, 812, L8
\bibitem[]{}Weingartner, J.C., \& Draine, B.T.\
                  2001, ApJ, 548, 296 
\bibitem[]{}Yang, L., Deslippe, J., Park, C.-H., 
                  Cohen, M.~L., \& Louie, S.~G.\ 2009, 
                  Phys. Rev. Lett., 103, 186802 
\bibitem[]{}Zhang, Y., \& Kwok, S.\ 2011, 
                 ApJ, 730, 126

\end{thebibliography}
\end{document}